\documentclass{aa}  
\usepackage[utf8]{inputenc}
\usepackage{threeparttable,pifont}
\usepackage{natbib}
\usepackage{color}
\usepackage{multicol}
\usepackage{amsmath,amssymb,url}
\usepackage{graphicx}
\usepackage{xcolor}
\usepackage{txfonts,longtable,dcolumn,verbatim}
\bibpunct{(}{)}{;}{a}{}{,}
\usepackage{array} 
\usepackage[normalem]{ulem}
\usepackage{lscape}
\usepackage{stfloats}   
\usepackage{placeins}   

\usepackage[squaren,textstyle]{SIunits}

\usepackage{hyperref}
\hypersetup{
    unicode=true,           
    pdftoolbar=true,        
    pdfmenubar=true,        
    pdffitwindow=true,      
    pdftitle={Physical models of Jupiter trojan asteroids},
    pdfauthor={Hanus et al.},
    pdfsubject={Planetary Science},
    pdfkeywords={},         
    pdfnewwindow=true,      
    colorlinks=true,        
    linkcolor=gray,         
    citecolor=blue,         
    filecolor=gray,         
    urlcolor=gray           
}

\usepackage{xspace}

\newcolumntype{d}{D{.}{.}{-1}}

\begin{document}

\title{Physical Characterization of Asteroid (16583) Oersted Combining Stellar Occultation and Photometric Data}
\author{
Josef Hanu\v{s}\inst{1} \and
Marco Delbo\inst{2,3} \and
Petr Pokorn\'y\inst{4,5,6} \and
Franck Marchis\inst{7,8} \and
Thomas M.~Esposito\inst{7,8}
}

\institute{
     Charles University, Faculty of Mathematics and Physics, Institute of Astronomy, V Hole\v sovi\v ck\'ach 2, CZ-18000, Prague 8, Czech Republic\label{auuk}\\
     \email{josef.hanus@mff.cuni.cz}
     \and
     Université Côte d’Azur, CNRS-Lagrange, Observatoire de la Côte d’Azur, CS 34229, 06304 Nice Cedex 4, France\label{oca}
     \and
     School of Physics and Astronomy, University of Leicester, Leicester LE1 7RH, UK\label{leicester}     
     \and
     Department of Physics, The Catholic University of America, Washington, DC 20064, USA\label{pok1}
     \and
     Astrophysics Science Division, NASA Goddard Space Flight Center, Greenbelt, MD 20771, USA\label{pok2}
     \and
     Center for Research and Exploration in Space Science and Technology, NASA/GSFC, Greenbelt, MD 20771, USA\label{pok3}
     \and 
     SETI Institute, Carl Sagan Center, 189 Bernado Avenue, Mountain View, CA 94043, USA\label{seti}
     \and
     Unistellar, 5 all\'e Marcel Leclerc, b\^atiment B, 13008 Marseille, France\label{uni}
}

\date{Received Month DD, YYYY; accepted Month DD, YYYY}
\titlerunning{Physical Characterization of Asteroid (16583) Oersted}

\abstract
{We report a successful observation of a stellar occultation by asteroid (16583) Oersted, enabling a detailed physical characterization of its shape, spin state, and surface properties.}
{Our goal is to determine the physical parameters of Oersted by combining multi-chord occultation timing, sparse optical photometry, and thermal infrared observations. Such asteroids (size$\sim$20 km) are rarely modeled in this detail due to observational limitations, making Oersted a valuable case study.}
{We applied convex lightcurve inversion to sparse photometric data to derive an initial shape and spin state. This model was then refined and scaled using non-convex shape modeling with the ADAM algorithm, incorporating constraints from the occultation chord profile. Thermophysical modeling based on WISE thermal infrared fluxes was used to determine the asteroid’s effective diameter, geometric albedo, and thermal inertia.}
{The non-convex shape model reveals localized surface concavities and provides a size estimate consistent with radiometric measurements. The derived thermal inertia is typical for asteroids of comparable size.}
{This work demonstrates the effectiveness of combining stellar occultations, photometry, and thermal infrared data for asteroid modeling and highlights the valuable contributions of citizen scientists, who played a key role in capturing the occultation and constraining the asteroid's profile.}

\keywords{asteroids -- methods: observational -- occultations -- astrometry -- citizen science}

\maketitle

\section{Introduction}

Asteroids are remnants of the early Solar System and preserve key information about its dynamical and compositional evolution. Physical characterization—including determination of shape, size, spin state, surface roughness, and thermal properties—is essential for understanding both their origin and evolution histories in the context of the overall asteroid population and its various sub-populations.

Various observational techniques have been developed to retrieve such asteroid physical parameters, including radar imaging \citep{Ostro2002}, direct spacecraft reconnaissance \citep{Thomas1996}, adaptive optics \citep{Viikinkoski2018}, thermal infrared observations \citet{Delbo2007a}, and photometric lightcurves \citep{Kaasalainen2002a}. Among ground-based methods, stellar occultations and lightcurve inversion stand out due to their accessibility, cost-efficiency, and complementarity. Disk-integrated photometric lightcurves are particularly effective at constraining the rotation period and general shape of an asteroid through convex inversion methods \citep{Kaasalainen2001a, Kaasalainen2001b}, while stellar occultations provide direct measurements of the asteroid's silhouette in the plane of the sky, enabling precise size estimation and detection of non-convex features \citep{Durech2011, Herald2020}. Therefore, these two methods complement each other, and together they allow for obtaining a size-calibrated non-convex model for, in principle, any observable asteroid.

A stellar occultation occurs when an asteroid passes in front of a star, momentarily blocking its light for observers located within a narrow ground track. The timing of these events from multiple stations can be used to reconstruct a two-dimensional projection of the asteroid’s shape at a known rotational phase. With the advent of high-precision star positions from Gaia astrometry \citep{Gaia2022}, predictions of occultation paths have become increasingly accurate, making coordinated multi-chord observations of a single asteroid during one night more feasible and scientifically productive.

Stellar occultation observations have long been the domain of dedicated citizen astronomers, who contribute valuable scientific data using their own equipment, time, and resources. The International Occultation Timing Association (IOTA) and the \texttt{Occult} software package\footnote{\url{http://www.lunar-occultations.com/iota/occult4.htm}} are central to this community, providing the coordination and tools necessary to plan, observe, and analyze occultation events. In recent years, a new group of contributors has emerged through the deployment of eVscope telescopes—smart, connected instruments operated by members of the Unistellar Network\footnote{\url{https://science.unistellar.com/}} \citep{Marchis2020}. This global network of citizen astronomers, coordinated by the SETI Institute, has begun contributing to stellar occultation campaigns. In 2024, the Unistellar Network recorded approximately 400 occultation observations—still a small fraction compared to the more than 4,000 observations reported by IOTA observers \citep{Herald2025pc}, but representing a growing complementary effort.

In this work, we present a comprehensive physical characterization of asteroid (16583) Oersted (hereafter referred to simply as Oersted), based on a multi-chord stellar occultation observed on March 3, 2024, sparse photometric data from various surveys, and thermal infrared fluxes from the WISE satellite \citep{Wright2010}. Oersted is a relatively small main-belt asteroid with a diameter of approximately 20~km. Due to their small angular extent, such asteroids cannot be resolved by adaptive optics or radar imaging, making indirect techniques critical. Our results contribute new constraints on the shape, spin axis, and thermal inertia in this underrepresented population, where reliable thermal inertia values are especially scarce.

Our methodology integrates convex lightcurve inversion, non-convex shape modeling using the ADAM algorithm \citep{Viikinkoski2015}, and thermophysical modeling based on the approach of \citet{Lagerros1996} as implemented in \citet{Delbo2004, Hanus2018b, Hung2022}. In particular, we demonstrate the value of combining stellar occultations and photometry to scale shape models, resolve spin-state ambiguities, and constrain surface properties such as thermal inertia and albedo. We also highlight the contribution of citizen scientists, who played a key role in capturing the occultation, and thus allowed constraining the asteroid’s profile.

\section{Data}

\subsection{Sparse photometry}\label{sec:lcs}

To derive the rotational properties of asteroid Oersted, we utilized relevant sparse-in-time photometric observations. These data were collected from several large-scale sky surveys, including the Catalina Sky Survey \citep[CSS;][]{Larson2003}, the Asteroid Terrestrial-impact Last Alert System \citep[ATLAS;][]{Tonry2018}, the All-Sky Automated Survey for Supernovae \citep[ASAS-SN;][]{Shappee2014, Kochanek2017, Hanus2021}, Gaia Data Release 3 \citep[GaiaDR3;][]{Tanga2023}, the Zwicky Transient Facility \citep[ZTF;][]{bellm2019}, and Pan-STARRS. The photometric points were retrieved from the Minor Planet Center and the AstDyS databases, and span multiple apparition epochs, providing sufficient coverage in aspect and illumination geometry for lightcurve inversion techniques. The data were processed using the pipeline as described in \citet{Hanus2021}. A detailed fit of these data to our best-fit convex shape model is shown in Appendix~\ref{appendix:occ}.

Although sparse photometry does not allow high-resolution lightcurve modeling for individual apparitions, it has been shown to be effective for constraining rotation periods and spin axis orientations when combined across multiple epochs \citep[e.g.,][]{Hanus2011}. In this study, the sparse data formed the basis for our initial convex inversion and provided the spin state solution used in subsequent non-convex modeling.

\subsection{Stellar occultation}

\begin{table*}
\scriptsize{
\centering
\caption{Summary of Oersted stellar occultation observations}
\label{tab:observers}
\begin{tabular}{llllllllllll}
\hline\hline
Observer & Location & Latitude & Longitude & Alt (m) & D Time (UTC) & D Type & D Err (s) & R Time (UTC) & R Type & R Err (s) \\
\hline
Karlis Suvcans & Lielauce, LV & +56 31 14.2 & +22 51 50.4 & 105 & 20:20:50.9 & M & & 20:20:50.9 & M & \\
Miroslav Pol\'a\v{c}ek & Plze\v{n}, CZ & +49 44 37.5 & +13 20 56.7 & 335 & 20:19:51.44 & D & 0.04 & 20:19:51.84 & R & 0.04 \\
Ji\v{r}\'i Pol\'ak & Plze\v{n}-Lhota, CZ & +49 41 41.1 & +13 19 15.6 & 339 & 20:19:50.589 & D & 0.010 & 20:19:51.695 & R & 0.010 \\
Ji\v{r}\'i Pol\'ak \& Michal Pol\'ak & Plze\v{n}-Lhota, CZ & +49 41 41.1 & +13 19 15.6 & 339 & 20:19:51.1 & D & 0.2 & 20:19:52.2 & R & 0.2 \\
Zdenek Sika\v{c} & Ruda, CZ & +50 08 19.8 & +13 51 50.7 & 450 & 20:19:54.524 & D & 0.013 & 20:19:56.034 & R & 0.013 \\
Luk\'a\v{s} Winkler & \'Utu\v{s}ice, CZ & +49 41 07.8 & +13 22 40.9 & 366 & 20:19:50.4 & D & 0.3 & 20:19:51.8 & R & 0.3 \\
Miroslav Pol\'a\v{c}ek & Plze\v{n}, CZ & +49 43 24.4 & +13 26 19.3 & 400 & 20:19:50.67 & D & 0.03 & 20:19:52.37 & R & 0.03 \\
Juris Sennikovs & Dobele, LV & +56 41 08.1 & +23 24 03.6 & 10 & 20:20:52.06 & D & 0.02 & 20:20:53.71 & R & 0.02 \\
Josef Hanu\v{s} & P\v{r}edenice, CZ & +49 37 28.1 & +13 23 02.5 & 378 & 20:19:49.87 & D & 0.05 & 20:19:51.37 & R & 0.05 \\
Michal Rottenborn & Tym\'akov, CZ & +49 43 28.9 & +13 31 43.8 & 421 & 20:19:50.836 & D & 0.015 & 20:19:52.311 & R & 0.015 \\
Karel Hal\'i\v{r} & Rokycany, CZ & +49 45 06.3 & +13 36 09.3 & 403 & 20:19:51.17 & D & 0.02 & 20:19:52.55 & R & 0.02 \\
Ji\v{r}\'i Kub\'anek & Hůrky, CZ & +49 44 38.5 & +13 40 27.1 & 440 & 20:19:51.387 & D & 0.008 & 20:19:52.642 & R & 0.008 \\
Ingvars Tomsons & Riga, LV & +56 53 50.0 & +24 04 30.0 & 13 & 20:20:55.56 & D & 0.11 & 20:20:56.3 & R & 0.2 \\
Ji\v{r}\'i Kub\'anek & Stra\v{s}ice, CZ & +49 44 35.7 & +13 44 56.3 & 538 & 20:19:51.95 & D & 0.03 & 20:19:52.10 & R & 0.03 \\
Ji\v{r}\'i Kub\'anek & Stra\v{s}ice, CZ & +49 44 35.7 & +13 44 56.3 & 538 & 20:19:52.20 & D & 0.02 & 20:19:52.298 & R & 0.015 \\
\hline
\end{tabular}
\tablefoot{The columns provide information on the observers and their recorded occultation data. The \textit{Observer} column lists the names of the observers, followed by their \textit{Location}, which includes the observing site (city and country). The \textit{Latitude} and \textit{Longitude} columns give the geographical coordinates of the observing station in degrees, while \textit{Alt (m)} represents the elevation of the site. The \textit{D Time (UTC)} and \textit{R Time (UTC)} columns indicate the recorded disappearance and reappearance times of the star due to the asteroid's shadow, respectively, in UTC. The \textit{D Type} and \textit{R Type} columns classify the disappearance and reappearance events, where ‘D’ and ‘R’ indicate detected occultation events, and ‘M’ denotes a miss (no occultation detected). The \textit{D Err (s)} and \textit{R Err (s)} columns provide the timing uncertainties for the disappearance and reappearance measurements in seconds. Data taken from the Occult database.}
}
\end{table*}

\begin{figure}
    \centering
    \includegraphics[width=0.99\linewidth, trim=0cm 0.1cm 0cm 0.0cm, clip]{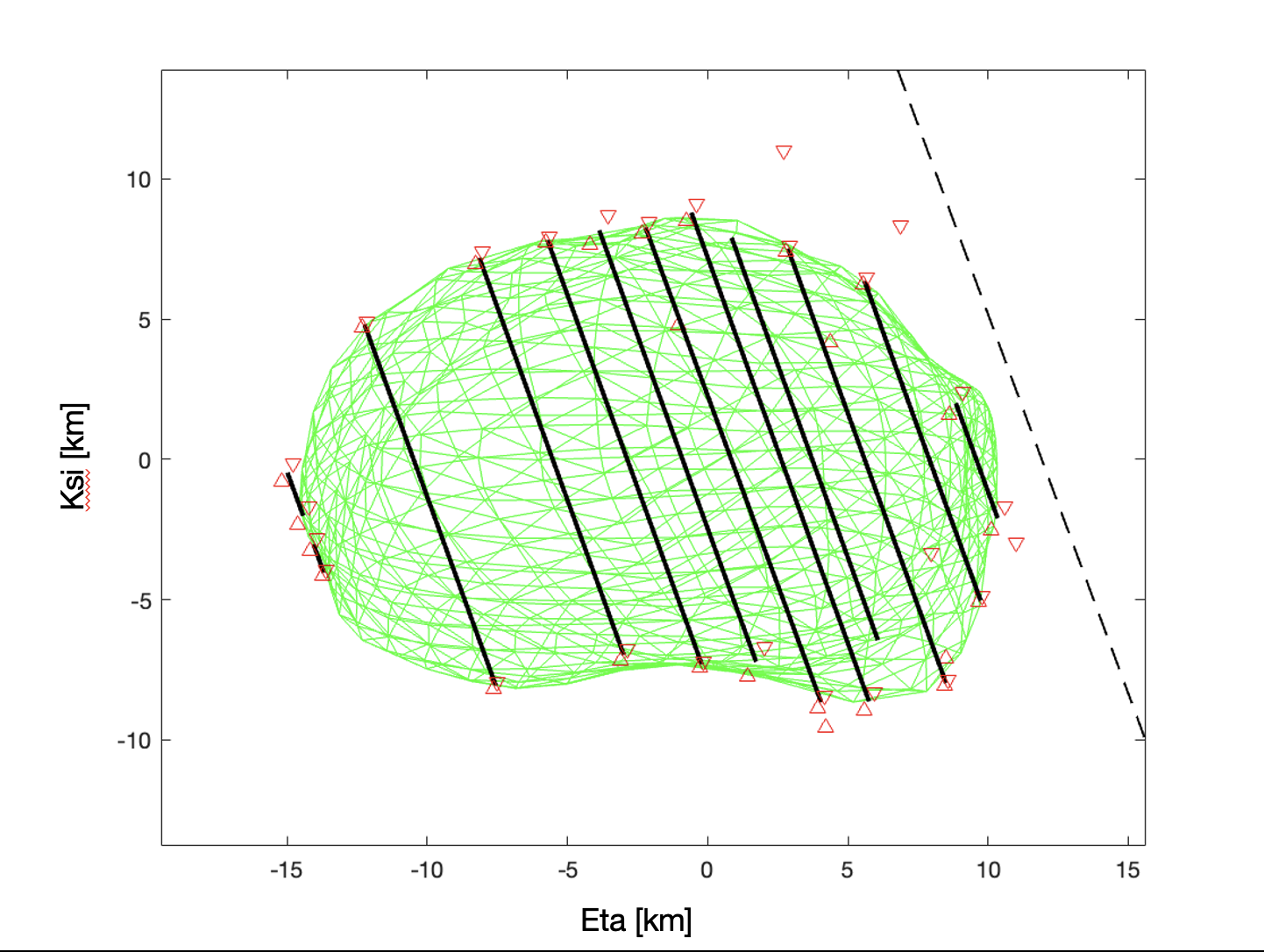}
    \caption{
    Observed occultation chords (black line segments) from the March 3, 2024 event, projected onto the sky plane, along with the best-fit non-convex ADAM shape model of asteroid Oersted (grey silhouette) at the corresponding rotational phase. Each chord represents the stellar disappearance and reappearance as seen from a specific observer location listed in Table~\ref{tab:observers}. The model silhouette is scaled and oriented to minimize residuals between the chord endpoints and the projected shape boundary. The close match demonstrates the validity of the shape and spin solution.}
    \label{fig:occultation_fit}
\end{figure}

A stellar occultation by asteroid Oersted occurred on March 3, 2024, involving a target star with apparent magnitude $V = 8.7$. The event was successfully recorded by fourteen observing stations located in the Czech Republic and Latvia, resulting in thirteen positive detections and one negative detection (miss). Two of the detections were obtained at the same observing site in Plze\v{n}-Lhota: one using video and one visually. The visual observation was included in the analysis with a manually applied time offset to align it with the video-derived timings (the \texttt{Occult} software already provides the chord with this offset). Additionally, the observation from Stra\v{s}ice shows a grazing occultation, where the star was briefly occulted twice (thus treated as two separate chords), likely due to local topographic features on the asteroid’s limb. Finally, the chord observed by L.~Winkler, with an uncertainty of 0.3s, is markedly shorter than the neighboring chords. We therefore excluded it from the shape modeling, as it likely adds no meaningful constraint relative to the other observations. This event provided a dense chord coverage across the asteroid's silhouette, making it one of the most successful stellar occultations of 2024.

The occultation data were downloaded in XML format from the \texttt{Occult} software maintained by D.~Herald. \texttt{Occult} is a widely used tool for predicting and reducing occultation events, providing standardized timing and geometry data. The occultation data archive is periodically backed up at NASA's Planetary Data System Small Bodies Node\footnote{\url{https://sbn.psi.edu/pds/resource/occ.html}} with the most recent version being \citep{Herald2024}. Table~\ref{tab:observers} summarizes the observing stations, recorded timings, and associated uncertainties. The occultation chords were extracted from individual video, CCD, or CMOS-based recordings using \texttt{Tangra}\footnote{\url{http://www.hristopavlov.net/Tangra/}}, \texttt{PyMovie}\footnote{\url{https://occultations.org/observing/software/pymovie/}}, or the Unistellar Network pipeline (observation by Josef Hanu\v{s}), and converted into a format compatible with shape modeling following the procedure of \citet{Durech2011}.

The predicted shadow path, based on Gaia DR3 stellar positions and the JPL\#54 ephemeris, was in excellent agreement with the observations. 

\subsection{Thermal infrared data}

Thermal infrared observations of Oersted were obtained from the WISE space telescope in the 12 and 22~$\mu$m bands \citep{Wright2010}. The data were downloaded from the IRSA/IPAC archive and filtered to include only high-quality detections with signal-to-noise ratio (S/N) greater than 5 \citep{Hanus2015a}. In total, five epochs include measurements in both filters, while two additional epochs contain data in only one filter. These thermal fluxes were used as input to our thermophysical modeling (TPM) to derive the asteroid's effective diameter, thermal inertia, and geometric visible albedo.

\section{Methods}

To characterize the physical properties of asteroid Oersted, we combined photometric lightcurve data with multi-chord stellar occultation observations. The analysis followed a three-step approach: convex lightcurve inversion, non-convex shape modeling using ADAM, and thermophysical modeling using WISE data.

\subsection{Convex lightcurve inversion}

\begin{figure}
    \centering
    \includegraphics[width=0.99\linewidth]{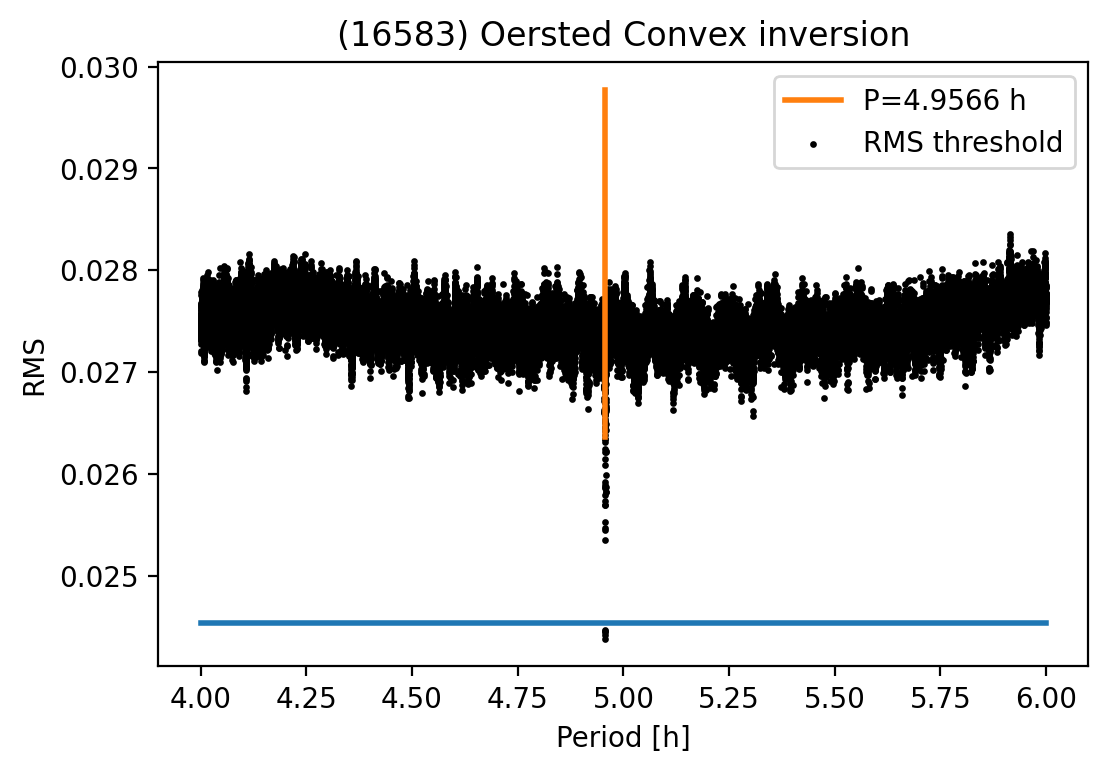}
    \caption{
    Periodogram of asteroid Oersted based on sparse photometric data. The global minimum, corresponding to a sidereal rotation period of $P_{\mathrm{sid}} = 4.956677$ hours, is indicated by the vertical line. The horizontal solid blue line marks the adopted threshold for the $\chi^2$ values used to evaluate the uniqueness of the period solutions.
}
    \label{fig:periodogram}
\end{figure}

We first applied the lightcurve inversion technique of \citet{Kaasalainen2001a,Kaasalainen2001b} to the available disk-integrated photometric data. The dataset includes sparse-in-time photometry from Gaia DR3, ASAS-SN, ATLAS, Pan-STARRS, ZTF, and Catalina Sky survey (see Sec.~\ref{sec:lcs} and Appendix~\ref{appendix:occ}).

The period search (Fig.~\ref{fig:periodogram}) reveals a well-defined global minimum corresponding to a sidereal rotation period of $P = 4.95663$ hours, representing the first determination of Oersted’s rotational period. While convex lightcurve inversion typically yields two mirror pole solutions due to geometric ambiguity, only a single pole solution was identified in this case (Table~\ref{table:rotation_states}). This uniqueness can likely be attributed to the relatively high orbital inclination of Oersted ($i = 22.6^\circ$), which reduces the pole degeneracy inherent in lightcurve inversion \citep{Kaasalainen2006}.

\subsection{Non-convex shape modeling with occultation constraints}

\begin{figure}
    \centering
    \includegraphics[width=0.33\hsize]{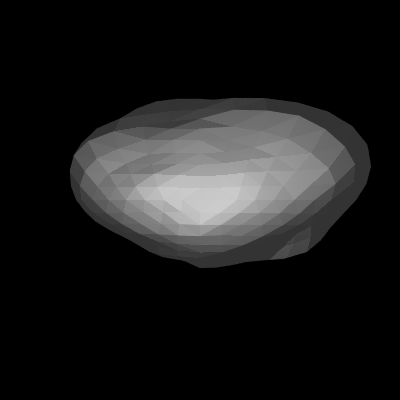}\includegraphics[width=0.33\hsize]{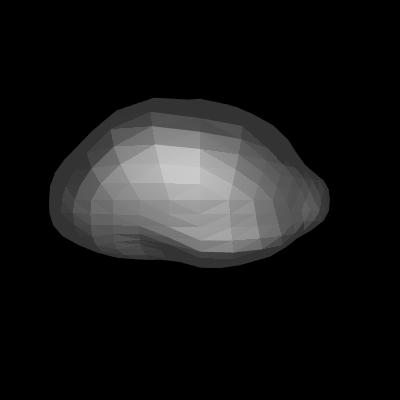}\includegraphics[width=0.33\hsize]{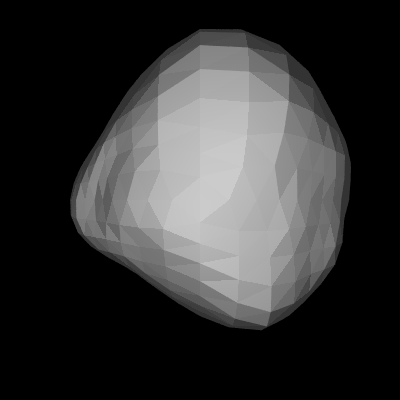}
    \caption{
    Non-convex shape model of asteroid Oersted reconstructed using the ADAM algorithm. The model was scaled to match the stellar occultation chords observed on March 3, 2024. The inclusion of occultation data enables the resolution of surface concavities and the overall elongation of the body. Three viewing perspectives are shown: the first two are equator-on views separated by $90^\circ$ in longitude, and the third is a pole-on view.
}
    \label{fig:shape_model}
\end{figure}

To refine the shape model, we employed the ADAM algorithm \citep{Viikinkoski2015}, which supports shape modeling with direct fitting to the occultation profile. We converted the occultation timings (Table~\ref{tab:observers}) into chords in the sky plane, following the procedure described by \citet{Durech2011}. The occultation was observed from 13 positive stations across Europe, providing strong constraints on the asteroid's shape projection.

The ADAM algorithm was initialized with the convex shape and spin parameters from the lightcurve inversion and iteratively refined to fit both the photometric and occultation data \citep[cf.][]{Marciniak2023, Viikinkoski2017}. We performed multiple ADAM runs using different initial meshes and regularization weights to assess the uncertainty in the derived size and topography. The fit to the occultation profile is shown in Fig.~\ref{fig:occultation_fit} and the resulting shape model in Fig.~\ref{fig:shape_model}.

\subsection{Thermophysical modeling}\label{sec:TPM}

To derive the thermophysical properties of Oersted, we employed the thermophysical model (TPM) originally developed by \citet{Lagerros1996} and later adapted by \citet{Delbo2004, Hanus2018b, Hung2022}. The TPM uses the asteroid's shape and spin state as input and fits observed thermal infrared fluxes, here specifically from the WISE mission at 12 and 22~$\mu$m bands \citep{Wright2010}. The model searches a grid of thermal inertia values ($\Gamma$), surface roughness parameters (mean slope angle $\bar{\theta}$), and geometric albedo ($p_\mathrm{V}$), minimizing the $\chi^2$ between synthetic and observed fluxes.

We investigated three thermophysical modeling configurations:

\begin{itemize}
    \item Convex model without color correction: The TPM was first applied to the convex shape derived from lightcurve inversion, with size as a free parameter. This model does not account for concavities or mutual heating effects but serves as a benchmark.

    \item Convex model with color correction: A second run used the same convex model but included the color correction for the thermal emissivity. The effect on the best-fit parameters was small, with a negligible impact on thermal inertia or fit quality compared to the run without color correction.

    \item Non-convex ADAM model (no color correction): We also applied the TPM to the non-convex ADAM shape model without including color correction. The current TPM implementation for non-convex geometries does not correctly handle this correction, and addressing the issue lies beyond the scope of this study. However, our comparison with the convex case shows that the effect of omitting color correction is negligible; therefore, we proceeded without it for the ADAM-based TPM.
\end{itemize}

All models assumed a Bond albedo of $A = 0.02$, derived from the volume-equivalent diameter obtained from ADAM modeling and the absolute magnitude $H = 12.28$ from the MPC. The geometric albedo $p_V$ was calculated using the standard relation $p_V = \left[\frac{1329}{D\,(\mathrm{km})} \times 10^{-H/5}\right]^2$, and the Bond albedo was then computed as $A = q\,p_V$, where the phase integral $q$ was estimated using a slope parameter of $G = 0.15$ (also from the MPC). The TPM analysis was performed over a grid of thermal inertia and surface roughness values. A summary of the resulting parameters is provided in Table~\ref{tab:tpm_results1}.

\begin{table*}[h]
\centering
\caption{Comparison of thermophysical properties of asteroid Oersted derived from three TPM configurations.}
\label{tab:tpm_results1}
\begin{tabular}{lccc}
\hline\hline
Parameter & Convex (no color corr.) & Convex (with color corr.) & ADAM (no color corr.) \\
\hline
Thermal inertia $\Gamma$ (J\,m$^{-2}$\,s$^{-1/2}$\,K$^{-1}$) & 37$^{+5}_{-12}$ & 37$^{+5}_{-12}$  & 37$^{+14}_{-12}$ \\
Mean surface slope $\bar{\theta}$ (deg) & 32.8/22.7 & 32.8/22.7 & 22.7--55.4 \\
Geometric albedo $p_V$ & 0.046 & 0.046 & 0.046 \\
Volume-equivalent diameter $D_\mathrm{eq}$ (km) & 21.6$\pm$0.6 & 21.8$\pm$0.6 & 20.9$\pm$0.6 \\
Reduced $\chi^2$ & 1.66 & 1.71 & 2.00 \\
\hline
\end{tabular}
\end{table*}

\section{Results}

\subsection{Spin state and shape}

\begin{table}[h]
\centering
\caption{Derived rotation states of asteroid Oersted from convex inversion and ADAM modeling.}
\begin{tabular}{lccc}
\hline\hline
Model & $\lambda$ (deg) & $\beta$ (deg) & $P$ (hours) \\
\hline
Convex inversion (CI) & $294 \pm 50$ & $+88 \pm 3$ & 4.95663 \\
ADAM (non-convex) & $244 \pm 30$ & $+86 \pm 2$ & 4.95665 \\
\hline
\end{tabular}
\label{table:rotation_states}
\end{table}

The convex lightcurve inversion and the ADAM modeling both converge on a consistent sidereal rotation period of $P = 4.95664 \pm 0.00001$ h. The derived pole solutions exhibit very similar ecliptic latitudes ($\beta$), while differing slightly in ecliptic longitude ($\lambda$). This discrepancy is not significant, as the pole latitude is close to $\beta \approx 90^\circ$, a region where small angular separations correspond to large changes in longitude due to the coordinate system geometry. As a result, the total angular separation between the solutions remains within 5$^\circ$, consistent with typical uncertainties in pole determination.

The resulting non-convex shape model (Fig.~\ref{fig:shape_model}) reveals localized surface concavities and a slightly elongated body with axis ratios of approximately 1.00:0.87:0.72. The volume-equivalent diameter derived from the ADAM model is $D = 21.7 \pm 1.0$ km, and the projected silhouette matches the occultation profile within the chord timing uncertainties (Fig.~\ref{fig:occultation_fit}).

The shape model, spin-state parameters, and optical photometric data are available from the Database of Asteroid Models from Inversion Techniques (DAMIT\footnote{\url{https://damit.cuni.cz/projects/damit/}}, \citealt{Durech2010}).

\subsection{Thermal properties}

\begin{figure*}
    \centering
    \includegraphics[width=0.48\linewidth]{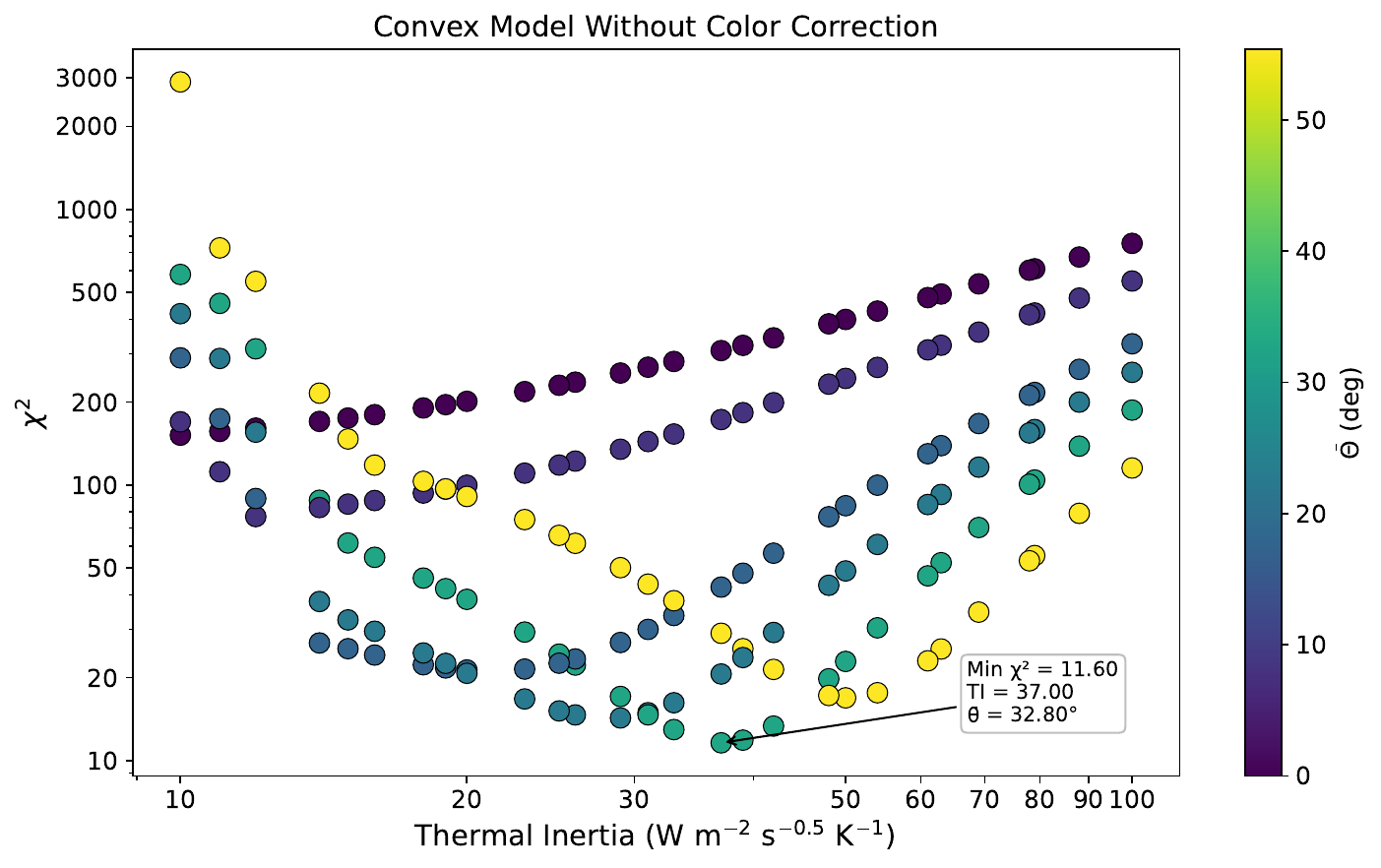}
    \includegraphics[width=0.48\linewidth]{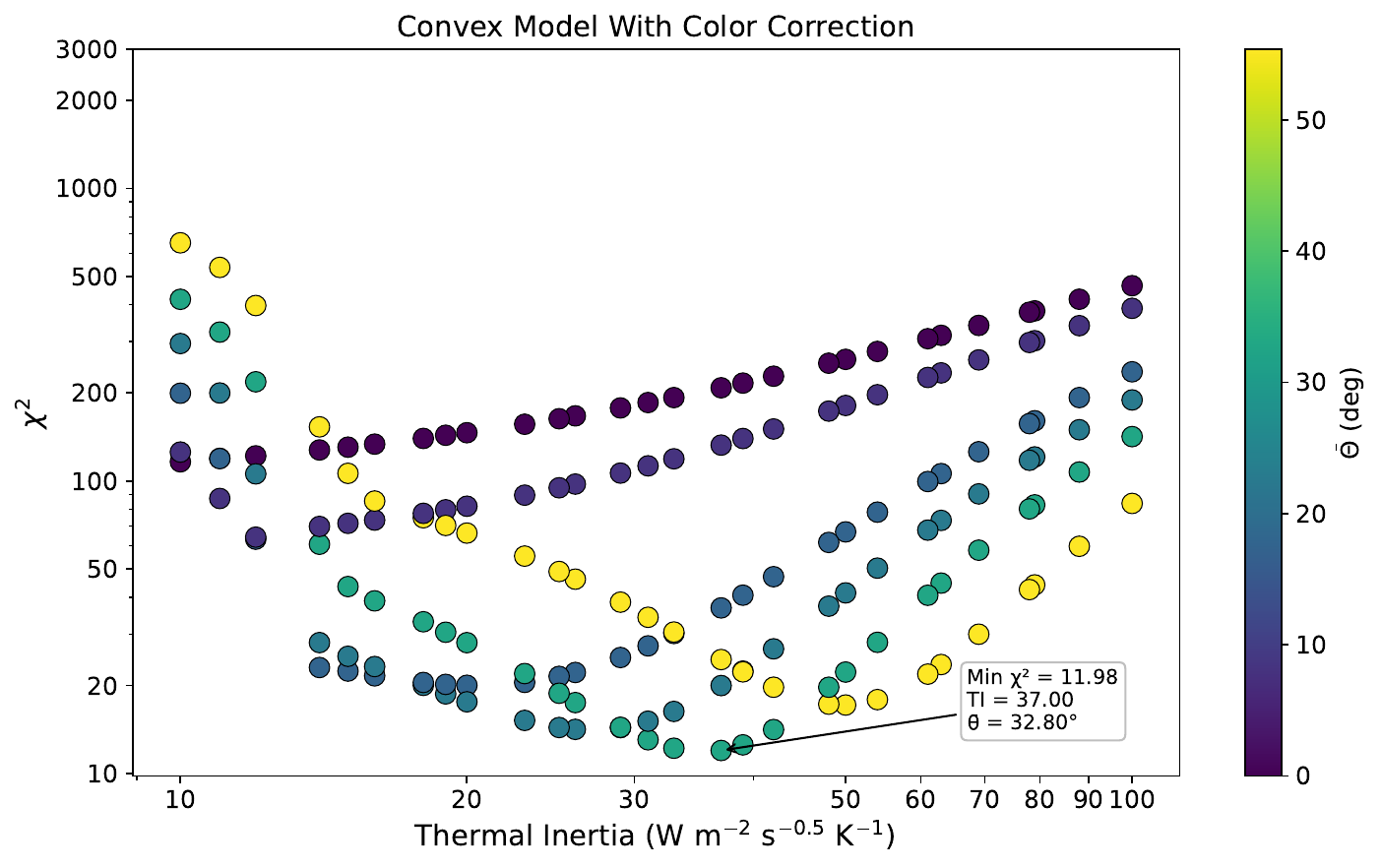}
    \caption{Comparison of thermophysical model fits using the convex shape of asteroid Oersted without (left) and with (right) color correction. The plot shows the chi-square ($\chi^2$) as a function of thermal inertia for various levels of surface roughness, characterized by the mean surface slope angle $\bar{\theta}$. Each curve corresponds to a different roughness value, as indicated in the legend. The best-fit thermal inertia is $\Gamma = 37$ J\,m$^{-2}$\,s$^{-1/2}$\,K$^{-1}$ in both cases. Color correction leads to a marginal decrease in the fit quality, increasing the minimum $\chi^2$ from 1.66 to 1.71, but does not significantly affect the derived physical parameters.}
    \label{fig:TPM_CC}
\end{figure*}

\begin{figure}
    \centering
    \includegraphics[width=\linewidth]{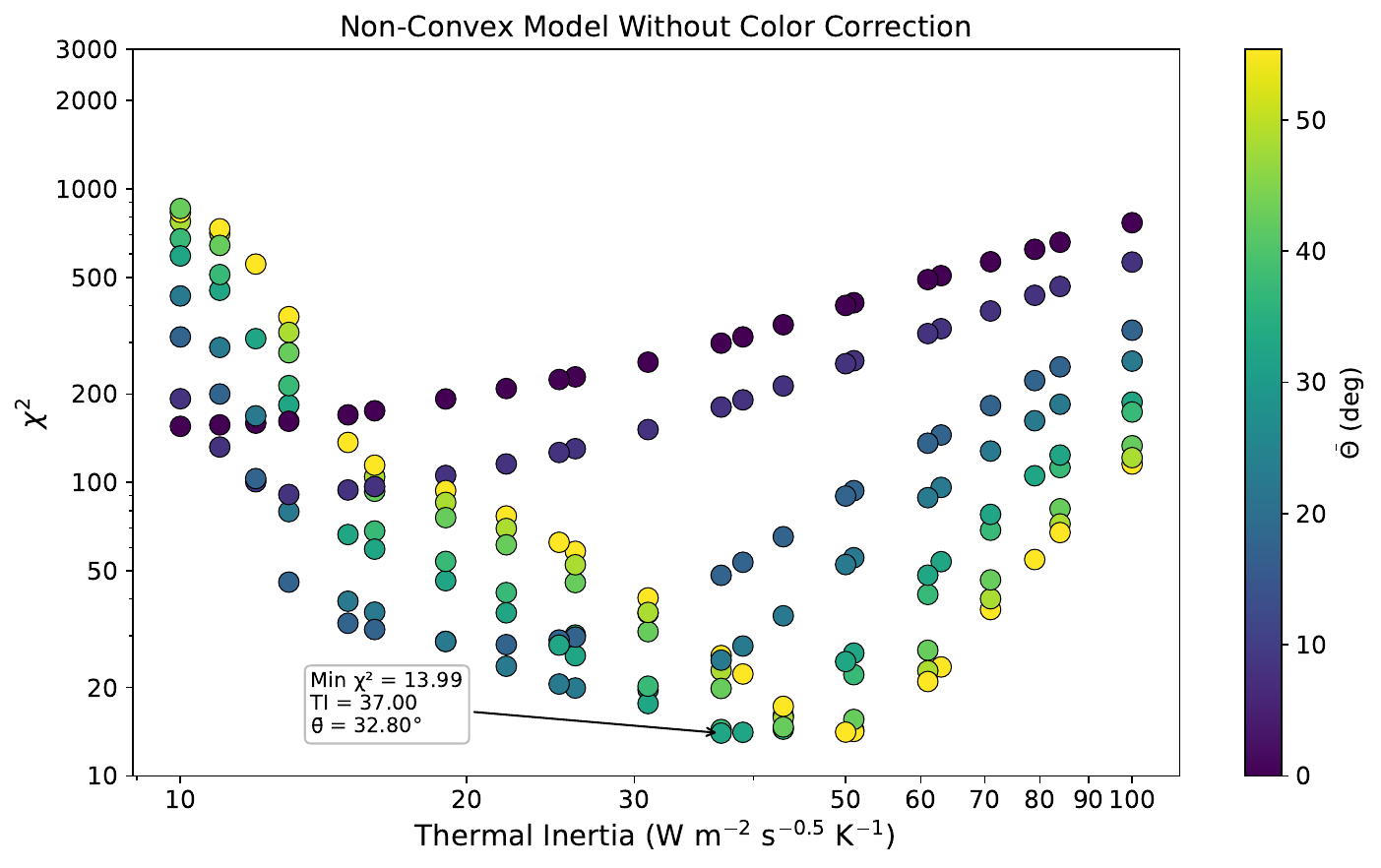}
    \caption{Thermophysical model fit for asteroid Oersted using the non-convex ADAM shape model.  The plot shows the chi-square ($\chi^2$) as a function of thermal inertia for various levels of surface roughness, characterized by the mean surface slope angle $\bar{\theta}$. Each curve corresponds to a different roughness value, as indicated in the legend. The best-fit solution corresponds to $\Gamma = 37$ J\,m$^{-2}$\,s$^{-1/2}$\,K$^{-1}$ and $\bar{\theta} = 32.8^\circ$, with a minimum reduced $\chi^2$ of 2.00.}
    \label{fig:TPM_ADAM}
\end{figure}

The thermophysical modeling results for asteroid Oersted are summarized in Table~\ref{tab:tpm_results1}. All configurations yield consistent sizes, with the volume-equivalent diameter ranging from 20.9 to 21.8 km and geometric visible albedo fixed at $p_V = 0.046$.

For the convex model without color correction, the best-fit thermal inertia is $\Gamma = 37$ J\,m$^{-2}$\,s$^{-1/2}$\,K$^{-1}$ and the mean surface slope is $\bar{\theta} = 32.8^\circ$, with a reduced $\chi^2$ of 1.66. Including the color correction increases the $\chi^2$ slightly to 1.71 but does not significantly affect the thermal parameters, indicating that color correction has only a minor impact (Oersted is a C-type asteroid with possibly a flat spectrum). We show the fit comparison in Fig~\ref{fig:TPM_CC}.

For the non-convex ADAM shape model (without color correction), the best-fit thermal inertia remains $\Gamma = 37$ J\,m$^{-2}$\,s$^{-1/2}$\,K$^{-1}$, as well as the roughness $\bar{\theta} = 32.8^\circ$. The fit quality slightly worsened, with a reduced $\chi^2$ of 2.00. Although color correction was not applied due to implementation limitations, a prior test using the convex model suggests that its effect is negligible.

These results confirm that the thermophysical properties derived are robust across modeling approaches, with thermal inertia values typical for main-belt asteroids of similar size \citep{Delbo2015}. The slightly elevated surface roughness inferred from the non-convex model is plausible, given the presence of concavities resolved in the ADAM shape.

\section{Discussion}

The non-convex model, as derived from the ADAM algorithm, shows localized topography, including potential concavities that could not be detected by convex methods alone. 

Our thermophysical modeling results are in excellent agreement with the radiometric estimates published by \citet{Masiero2011}, who derived a volume-equivalent diameter of $22.10 \pm 0.22$ km and a geometric visible albedo of $p_V = 0.045$ for Oersted using NEOWISE thermal infrared data and a standard thermal model. However, the quoted relative uncertainty of $\sim$1\% is unrealistically low, as noted by \citet{Mainzer2016} and further supported by \citet{Herald2020}, who compared radiometric diameters with independent stellar occultation measurements for 219 asteroids. Therefore, in practice, radiometric diameters typically carry uncertainties of at least 10\% due to limitations in shape, thermal properties, and observation geometry. The values derived from our convex and non-convex shape models ($D = 20.9$–$21.8$ km and $p_V = 0.046$) are fully consistent with the NEOWISE estimates, providing an important cross-validation of both modeling approaches. The close match supports the robustness of our shape models and confirms that even when self-heating effects are incorporated, the derived size and albedo remain stable. This agreement strengthens confidence in the use of occultation-constrained shape modeling combined with thermophysical analysis as a reliable alternative to classical radiometric methods.

The thermophysical properties derived for Oersted, in particular the thermal inertia $\Gamma = 37^{+14}_{-12}$ J m$^{-2}$ s$^{-1/2}$ K$^{-1}$ (we adopt the solution with the non-convex shape), are consistent with values typically found among similarly sized main-belt asteroids \citep{Delbo2015,Hung2022,Novakovic2024}. This moderate thermal inertia suggests that the surface is likely covered by a layer of coarse-grained or compact regolith, rather than fine dust (TI<20) or bare rock (TI>500). Such a surface structure would allow for partial heat retention between day-night cycles, but without the extremely low conductivity associated with very fine-grained regolith or the high conductivity of exposed rock.

The agreement in effective diameter between the thermophysical model and the occultation-constrained ADAM shape confirms the reliability of both approaches. While the flux color correction was not included in the non-convex TPM due to implementation limitations, the usage of the ADAM model still provides a more physically realistic surface representation. The consistency in diameter and fit quality across modeling setups highlights the importance of including detailed shape information when interpreting thermal infrared fluxes, especially for asteroids with resolved topographic features.

These results further demonstrate the synergy between lightcurve inversion, stellar occultations, and thermal infrared data in producing physically realistic asteroid models. For objects like Oersted, where multi-chord occultations and quality photometry are available, the combination of methods enables not only accurate shape reconstruction but also meaningful constraints on surface composition and thermal behavior.

\section{Conclusions}

We have presented a comprehensive physical characterization of asteroid Oersted based on multi-chord stellar occultation data, disk-integrated photometry, and thermal infrared observations (see Tables~\ref{tab:tpm_results1} and \ref{table:rotation_states}). Asteroids in this size range ($D\sim20$ km) are rarely modeled in this detail due to observational limitations, making Oersted a valuable case study. 

Our analysis combined convex lightcurve inversion, non-convex shape modeling using the ADAM algorithm, and thermophysical modeling constrained by WISE data. The resulting shape model resolves surface concavities, while the derived size and thermal properties are consistent with expectations for a mid-sized main-belt asteroid with a moderately compact regolith. 

Crucially, this study demonstrates the power of combining complementary datasets—particularly occultations and lightcurves—to achieve high-fidelity asteroid models. The successful inclusion of data from the citizen astronomers underscores the growing potential of distributed observing campaigns to contribute valuable constraints in planetary science and further highlights the role of community-driven efforts in advancing our understanding of small bodies in the Solar System.

Multi-chord occultation observations remain a key tool for constraining asteroid shapes and sizes. The growing availability of automated telescopes, such as those in the Unistellar Network, offers a promising opportunity to increase the frequency and spatial coverage of such events. Future efforts should focus on better integrating this existing infrastructure into a broader, decentralized global network, enabling more efficient coordination and data sharing for small body characterization \citep{Marchis2025}.

\begin{acknowledgements}
We gratefully acknowledge all amateur and professional astronomers who recorded and submitted occultation observations used in this study. Their commitment of time and resources is essential for advancing asteroid science. We also thank Dave Herald for his thorough review and constructive suggestions, which helped improve the quality and clarity of the manuscript. The grant 22-17783S of the Czech Science Foundation has supported this work.
\end{acknowledgements}

\bibliographystyle{aa}
\bibliography{mybib}

\begin{thebibliography}{37}
\expandafter\ifx\csname natexlab\endcsname\relax\def\natexlab#1{#1}\fi

\bibitem[{{Bellm} {et~al.}(2019){Bellm}, {Kulkarni}, {Graham}, {Dekany}, {Smith}, {Riddle}, {Masci}, {Helou}, {Prince}, {Adams}, {Barbarino}, {Barlow}, {Bauer}, {Beck}, {Belicki}, {Biswas}, {Blagorodnova}, {Bodewits}, {Bolin}, {Brinnel}, {Brooke}, {Bue}, {Bulla}, {Burruss}, {Cenko}, {Chang}, {Connolly}, {Coughlin}, {Cromer}, {Cunningham}, {De}, {Delacroix}, {Desai}, {Duev}, {Eadie}, {Farnham}, {Feeney}, {Feindt}, {Flynn}, {Franckowiak}, {Frederick}, {Fremling}, {Gal-Yam}, {Gezari}, {Giomi}, {Goldstein}, {Golkhou}, {Goobar}, {Groom}, {Hacopians}, {Hale}, {Henning}, {Ho}, {Hover}, {Howell}, {Hung}, {Huppenkothen}, {Imel}, {Ip}, {Ivezi{\'c}}, {Jackson}, {Jones}, {Juric}, {Kasliwal}, {Kaspi}, {Kaye}, {Kelley}, {Kowalski}, {Kramer}, {Kupfer}, {Landry}, {Laher}, {Lee}, {Lin}, {Lin}, {Lunnan}, {Giomi}, {Mahabal}, {Mao}, {Miller}, {Monkewitz}, {Murphy}, {Ngeow}, {Nordin}, {Nugent}, {Ofek}, {Patterson}, {Penprase}, {Porter}, {Rauch}, {Rebbapragada}, {Reiley}, {Rigault}, {Rodriguez}, {van Roestel}, {Rusholme}, {van
  Santen}, {Schulze}, {Shupe}, {Singer}, {Soumagnac}, {Stein}, {Surace}, {Sollerman}, {Szkody}, {Taddia}, {Terek}, {Van Sistine}, {van Velzen}, {Vestrand}, {Walters}, {Ward}, {Ye}, {Yu}, {Yan}, \& {Zolkower}}]{bellm2019}
{Bellm}, E.~C., {Kulkarni}, S.~R., {Graham}, M.~J., {et~al.} 2019, \pasp, 131, 018002

\bibitem[{Delbo'(2004)}]{Delbo2004}
Delbo', M. 2004, PhD thesis - Freie Univesitaet Berlin, 1

\bibitem[{{Delbo'} {et~al.}(2007){Delbo'}, {dell'Oro}, {Harris}, {Mottola}, \& {Mueller}}]{Delbo2007a}
{Delbo'}, M., {dell'Oro}, A., {Harris}, A.~W., {Mottola}, S., \& {Mueller}, M. 2007, \icarus, 190, 236

\bibitem[{{Delbo'} {et~al.}(2015){Delbo'}, {Mueller}, {Emery}, {Rozitis}, \& {Capria}}]{Delbo2015}
{Delbo'}, M., {Mueller}, M., {Emery}, J., {Rozitis}, B., \& {Capria}, M.~T. 2015, in Asteroids IV, ed. P.~{Michel}, F.~E. {DeMeo}, \& W.~F. {Bottke} (The University of Arizona Press), 107--128

\bibitem[{{\v{D}urech} {et~al.}(2011){\v{D}urech}, {Kaasalainen}, {Herald}, {Dunham}, {Timerson}, {Hanu\v{s}}, {Frappa}, {Talbot}, {Hayamizu}, {Warner}, {Pilcher}, \& {Gal{\'a}d}}]{Durech2011}
{\v{D}urech}, J., {Kaasalainen}, M., {Herald}, D., {et~al.} 2011, Icarus, 214, 652

\bibitem[{{\v Durech} {et~al.}(2010){\v Durech}, {Sidorin}, \& {Kaasalainen}}]{Durech2010}
{\v Durech}, J., {Sidorin}, V., \& {Kaasalainen}, M. 2010, \aap, 513, A46

\bibitem[{{Gaia Collaboration} {et~al.}(2022){Gaia Collaboration}, Vallenari, Brown, Prusti, de~Bruijne, Babusiaux, Tanga, \& et~al.}]{Gaia2022}
{Gaia Collaboration}, Vallenari, A., Brown, A.~G.~A., {et~al.} 2022, A\&A, 667, A1

\bibitem[{{Hanu\v{s}} {et~al.}(2015){Hanu\v{s}}, {Delbo'}, {\v Durech}, \& {Al{\'{\i}}-Lagoa}}]{Hanus2015a}
{Hanu\v{s}}, J., {Delbo'}, M., {\v Durech}, J., \& {Al{\'{\i}}-Lagoa}, V. 2015, \icarus, 256, 101

\bibitem[{{Hanu\v{s}} {et~al.}(2018){Hanu\v{s}}, {Delbo'}, {\v{D}urech}, \& {Al{\'{\i}}-Lagoa}}]{Hanus2018b}
{Hanu\v{s}}, J., {Delbo'}, M., {\v{D}urech}, J., \& {Al{\'{\i}}-Lagoa}, V. 2018, \icarus, 309, 297

\bibitem[{{Hanu\v{s}} {et~al.}(2011){Hanu\v{s}}, {\v{D}urech}, {Bro\v{z}}, {Warner}, {Pilcher}, {Stephens}, {Oey}, {Bernasconi}, {Casulli}, {Behrend}, {Polishook}, {Henych}, {Lehk{\'y}}, {Yoshida}, \& {Ito}}]{Hanus2011}
{Hanu\v{s}}, J., {\v{D}urech}, J., {Bro\v{z}}, M., {et~al.} 2011, \aap, 530, A134

\bibitem[{{Hanu{\v{s}}} {et~al.}(2021){Hanu{\v{s}}}, {Pejcha}, {Shappee}, {Kochanek}, {Stanek}, \& {Holoien}}]{Hanus2021}
{Hanu{\v{s}}}, J., {Pejcha}, O., {Shappee}, B.~J., {et~al.} 2021, \aap, 654, A48

\bibitem[{Herald(2025)}]{Herald2025pc}
Herald, D. 2025, personal communication

\bibitem[{{Herald} {et~al.}(2020){Herald}, {Gault}, {Anderson}, {Dunham}, {Frappa}, {Hayamizu}, {Kerr}, {Miyashita}, {Moore}, {Pavlov}, {Preston}, {Talbot}, \& {Timerson}}]{Herald2020}
{Herald}, D., {Gault}, D., {Anderson}, R., {et~al.} 2020, \mnras, 499, 4570

\bibitem[{{Herald} {et~al.}(2024){Herald}, {Gault}, {Carlson}, {Guhl}, {Frappa}, {Giacchini}, {Hayamizu}, {Kerr}, \& {Moore}}]{Herald2024}
{Herald}, D., {Gault}, D., {Carlson}, N., {et~al.} 2024, {Small Bodies Occultations Bundle V4.0}, NASA Planetary Data System, urn:nasa:pds:smallbodiesoccultations::4.0

\bibitem[{{Hung} {et~al.}(2022){Hung}, {Hanu{\v{s}}}, {Masiero}, \& {Tholen}}]{Hung2022}
{Hung}, D., {Hanu{\v{s}}}, J., {Masiero}, J.~R., \& {Tholen}, D.~J. 2022, \psj, 3, 56

\bibitem[{{Kaasalainen} \& {Lamberg}(2006)}]{Kaasalainen2006}
{Kaasalainen}, M. \& {Lamberg}, L. 2006, Inverse Problems, 22, 749

\bibitem[{{Kaasalainen} {et~al.}(2002){Kaasalainen}, {Mottola}, \& {Fulchignoni}}]{Kaasalainen2002a}
{Kaasalainen}, M., {Mottola}, S., \& {Fulchignoni}, M. 2002, Asteroids III, 139

\bibitem[{{Kaasalainen} \& {Torppa}(2001)}]{Kaasalainen2001a}
{Kaasalainen}, M. \& {Torppa}, J. 2001, Icarus, 153, 24

\bibitem[{{Kaasalainen} {et~al.}(2001){Kaasalainen}, {Torppa}, \& {Muinonen}}]{Kaasalainen2001b}
{Kaasalainen}, M., {Torppa}, J., \& {Muinonen}, K. 2001, Icarus, 153, 37

\bibitem[{{Kochanek} {et~al.}(2017){Kochanek}, {Shappee}, {Stanek}, {Holoien}, {Thompson}, {Prieto}, {Dong}, {Shields}, {Will}, \& {Britt}}]{Kochanek2017}
{Kochanek}, C.~S., {Shappee}, B.~J., {Stanek}, K.~Z., {et~al.} 2017, \pasp, 129, 104502

\bibitem[{{Lagerros}(1996)}]{Lagerros1996}
{Lagerros}, J.~S.~V. 1996, \aap, 310, 1011

\bibitem[{{Larson} {et~al.}(2003){Larson}, {Beshore}, {Hill}, {Christensen}, {McLean}, {Kolar}, {McNaught}, \& {Garradd}}]{Larson2003}
{Larson}, S., {Beshore}, E., {Hill}, R., {et~al.} 2003, in {Bulletin of the American Astronomical Society}, Vol.~35, {AAS/Division for Planetary Sciences Meeting Abstracts \#35}, 982

\bibitem[{{Mainzer} {et~al.}(2016){Mainzer}, {Bauer}, {Cutri}, {Grav}, {Kramer}, {Masiero}, {Nugent}, {Sonnett}, {Stevenson}, \& {Wright}}]{Mainzer2016}
{Mainzer}, A.~K., {Bauer}, J.~M., {Cutri}, R.~M., {et~al.} 2016, NASA Planetary Data System, 247

\bibitem[{Marchis(2020)}]{Marchis2020}
Marchis, F. 2020, Astronomical Journal, 159, 123

\bibitem[{Marchis {et~al.}(2025)Marchis, Esposito, Hanu{\v{s}}, Pokorn{\'y}, Lambert, Pilorz, Graykowski, \& Sgro}]{Marchis2025}
Marchis, F., Esposito, T.~M., Hanu{\v{s}}, J., {et~al.} 2025, in EPSC-DPS Joint Meeting 2025, Helsinki, Finland, ePSC-DPS2025-1038

\bibitem[{{Marciniak} {et~al.}(2023){Marciniak}, {{\v{D}}urech}, {Choukroun}, {Hanu{\v{s}}}, {Og{\l}oza}, {Szak{\'a}ts}, {Moln{\'a}r}, {P{\'a}l}, {Monteiro}, {Frappa}, {Beisker}, {Pavlov}, {Moore}, {Adomavi{\v{c}}ien{\.{e}}}, {Aikawa}, {Andersson}, {Antonini}, {Argentin}, {Asai}, {Assoignon}, {Barton}, {Baruffetti}, {Bath}, {Behrend}, {Benedyktowicz}, {Bernasconi}, {Biguet}, {Billiani}, {B{\l}a{\.z}ewicz}, {Boninsegna}, {Borkowski}, {Bosch}, {Brazill}, {Bronikowska}, {Bruno}, {Butkiewicz-B{\k{a}}k}, {Caron}, {Casalnuovo}, {Castellani}, {Ceravolo}, {Conjat}, {Delincak}, {Delpau}, {Demeautis}, {Demirkol}, {Dr{\'o}{\.z}d{\.z}}, {Duffard}, {Durandet}, {Eisfeldt}, {Evangelista}, {Fauvaud}, {Fauvaud}, {Ferrais}, {Filipek}, {Fini}, {Fukui}, {G{\"a}hrken}, {Geier}, {George}, {Goffin}, {Golonka}, {Goto}, {Grice}, {Guhl}, {Hal{\'\i}{\v{r}}}, {Hanna}, {Harman}, {Hashimoto}, {Hasubick}, {Higgins}, {Higuchi}, {Hirose}, {Hirsch}, {Hofschulz}, {Horaguchi}, {Horbowicz}, {Ida}, {Ign{\'a}cz}, {Ishida}, {Isobe}, {Jehin},
  {Joachimczyk}, {Jones}, {Juan}, {Kami{\'n}ski}, {Kami{\'n}ska}, {Kankiewicz}, {Kasebe}, {Kattentidt}, {Kim}, {Kim}, {Kitazaki}, {Klotz}, {Komraus}, {Konstanciak}, {K{\"o}nyves-T{\'o}th}, {Kouno}, {Kowald}, {Krajewski}, {Krannich}, {Kreutzer}, {Kryszczy{\'n}ska}, {Kub{\'a}nek}, {Kudak}, {Kugel}, {Kukita}, {Kulczak}, {Lazzaro}, {Licandro}, {Livet}, {Maley}, {Manago}, {M{\'a}nek}, {Manna}, {Matsushita}, {Meister}, {Mesquita}, {Messner}, {Michelet}, {Michimani}, {Mieczkowska}, {Morales}, {Motyli{\'n}ski}, {Murawiecka}, {Newman}, {Nikitin}, {Nishimura}, {Oey}, {Oszkiewicz}, {Owada}, {Pak{\v{s}}tien{\.{e}}}, {Paw{\l}owski}, {Pereira}, {Perig}, {Per{\l}a}, {Pilcher}, {Podlewska-Gaca}, {Pol{\'a}k}, {Polakis}, {Poli{\'n}ska}, {Popowicz}, {Richard}, {Rives}, {Rodrigues}, {Rogi{\'n}ski}, {Rond{\'o}n}, {Rottenborn}, {Sch{\"a}fer}, {Schnabel}, {Schreurs}, {Selva}, {Simon}, {Skiff}, {Skrutskie}, {Skrzypek}, {Sobkowiak}, {Sonbas}, {Sposetti}, {Stuart}, {Szyszka}, {Terakubo}, {Thomas}, {Trela}, {Uchiyama}, {Urbanik},
  {Vaudescal}, {Venable}, {Watanabe}, {Watanabe}, {Winiarski}, {Wr{\'o}blewski}, {Yamamura}, {Yamashita}, {Yoshihara}, {Zawilski}, {Zelen{\'y}}, {{\.Z}ejmo}, {{\.Z}ukowski}, \& {{\.Z}ywica}}]{Marciniak2023}
{Marciniak}, A., {{\v{D}}urech}, J., {Choukroun}, A., {et~al.} 2023, \aap, 679, A60

\bibitem[{{Masiero} {et~al.}(2011){Masiero}, {Mainzer}, {Grav}, {Bauer}, {Cutri}, {Dailey}, {Eisenhardt}, {McMillan}, {Spahr}, {Skrutskie}, {Tholen}, {Walker}, {Wright}, {DeBaun}, {Elsbury}, {Gautier}, {Gomillion}, \& {Wilkins}}]{Masiero2011}
{Masiero}, J.~R., {Mainzer}, A.~K., {Grav}, T., {et~al.} 2011, \apj, 741, 68

\bibitem[{{Novakovi{\'c}} {et~al.}(2024){Novakovi{\'c}}, {Fenucci}, {Mar{\v{c}}eta}, \& {Pavela}}]{Novakovic2024}
{Novakovi{\'c}}, B., {Fenucci}, M., {Mar{\v{c}}eta}, D., \& {Pavela}, D. 2024, \psj, 5, 11

\bibitem[{{Ostro} {et~al.}(2002){Ostro}, {Hudson}, {Benner}, {Giorgini}, {Magri}, {Margot}, \& {Nolan}}]{Ostro2002}
{Ostro}, S.~J., {Hudson}, R.~S., {Benner}, L.~A.~M., {et~al.} 2002, Asteroids III, 151

\bibitem[{{Shappee} {et~al.}(2014){Shappee}, {Prieto}, {Grupe}, {Kochanek}, {Stanek}, {De Rosa}, {Mathur}, {Zu}, {Peterson}, \& {Pogge}}]{Shappee2014}
{Shappee}, B.~J., {Prieto}, J.~L., {Grupe}, D., {et~al.} 2014, \apj, 788, 48

\bibitem[{{Tanga} {et~al.}(2023){Tanga}, {Pauwels}, {Mignard}, {Muinonen}, {Cellino}, {David}, {Hestroffer}, {Spoto}, {Berthier}, {Guiraud}, {Roux}, {Carry}, {Delbo}, {Dell'Oro}, {Fouron}, {Galluccio}, {Jonckheere}, {Klioner}, {Lefustec}, {Liberato}, {Ord{\'e}novic}, {Oreshina-Slezak}, {Penttil{\"a}}, {Pailler}, {Panem}, {Petit}, {Portell}, {Poujoulet}, {Thuillot}, {Van Hemelryck}, {Burlacu}, {Lasne}, \& {Managau}}]{Tanga2023}
{Tanga}, P., {Pauwels}, T., {Mignard}, F., {et~al.} 2023, \aap, 674, A12

\bibitem[{{Thomas} {et~al.}(1996){Thomas}, {Belton}, {Carcich}, {Chapman}, {Davies}, {Sullivan}, \& {Veverka}}]{Thomas1996}
{Thomas}, P.~C., {Belton}, M.~J.~S., {Carcich}, B., {et~al.} 1996, Icarus, 120, 20

\bibitem[{{Tonry} {et~al.}(2018){Tonry}, {Denneau}, {Heinze}, {Stalder}, {Smith}, {Smartt}, {Stubbs}, {Weiland}, \& {Rest}}]{Tonry2018}
{Tonry}, J.~L., {Denneau}, L., {Heinze}, A.~N., {et~al.} 2018, \pasp, 130, 064505

\bibitem[{{Viikinkoski} {et~al.}(2017){Viikinkoski}, {Hanu{\v s}}, {Kaasalainen}, {Marchis}, \& {{\v D}urech}}]{Viikinkoski2017}
{Viikinkoski}, M., {Hanu{\v s}}, J., {Kaasalainen}, M., {Marchis}, F., \& {{\v D}urech}, J. 2017, \aap, 607, A117

\bibitem[{{Viikinkoski} {et~al.}(2015){Viikinkoski}, {Kaasalainen}, \& {\v{D}urech}}]{Viikinkoski2015}
{Viikinkoski}, M., {Kaasalainen}, M., \& {\v{D}urech}, J. 2015, \aap, 576, A8

\bibitem[{{Viikinkoski} {et~al.}(2018){Viikinkoski}, {Vernazza}, {Hanu{\v s}}, {Le Coroller}, {Tazhenova}, {Carry}, {Marsset}, {Drouard}, {Marchis}, {Fetick}, {Fusco}, {{\v D}urech}, {Birlan}, {Berthier}, {Bartczak}, {Dumas}, {Castillo-Rogez}, {Cipriani}, {Colas}, {Ferrais}, {Grice}, {Jehin}, {Jorda}, {Kaasalainen}, {Kryszczynska}, {Lamy}, {Marciniak}, {Michalowski}, {Michel}, {Pajuelo}, {Podlewska-Gaca}, {Santana-Ros}, {Tanga}, {Vachier}, {Vigan}, {Warner}, {Witasse}, \& {Yang}}]{Viikinkoski2018}
{Viikinkoski}, M., {Vernazza}, P., {Hanu{\v s}}, J., {et~al.} 2018, \aap, 619, L3

\bibitem[{{Wright} {et~al.}(2010){Wright}, {Eisenhardt}, {Mainzer}, {Ressler}, {Cutri}, {Jarrett}, {Kirkpatrick}, {Padgett}, {McMillan}, {Skrutskie}, {Stanford}, {Cohen}, {Walker}, {Mather}, {Leisawitz}, {Gautier}, {McLean}, {Benford}, {Lonsdale}, {Blain}, {Mendez}, {Irace}, {Duval}, {Liu}, {Royer}, {Heinrichsen}, {Howard}, {Shannon}, {Kendall}, {Walsh}, {Larsen}, {Cardon}, {Schick}, {Schwalm}, {Abid}, {Fabinsky}, {Naes}, \& {Tsai}}]{Wright2010}
{Wright}, E.~L., {Eisenhardt}, P.~R.~M., {Mainzer}, A.~K., {et~al.} 2010, \aj, 140, 1868

\end{thebibliography}

\newpage
\begin{appendix}
\section{Additional figures}\label{appendix:occ}

To complement the main analysis presented in this work, we include in this appendix a series of diagnostic plots illustrating the quality of the photometric fits. These figures show the comparison between observed and modeled lightcurves for all sparse photometric datasets used in the shape and spin modeling of asteroid Oersted. The datasets include measurements from Gaia DR3, ASAS-SN, ATLAS, Pan-STARRS, ZTF, and the Catalina Sky Survey. Each figure consists of three panels per dataset, visualizing (i) the phase-folded lightcurve and corresponding model, (ii) residuals relative to the fitted phase function, and (iii) residuals relative to the normalized lightcurve. These plots serve to demonstrate the internal consistency of the lightcurve inversion results and the reliability of the shape model.

\FloatBarrier

\begin{figure*}[!b]
    \centering
    \includegraphics[width=0.81\linewidth]{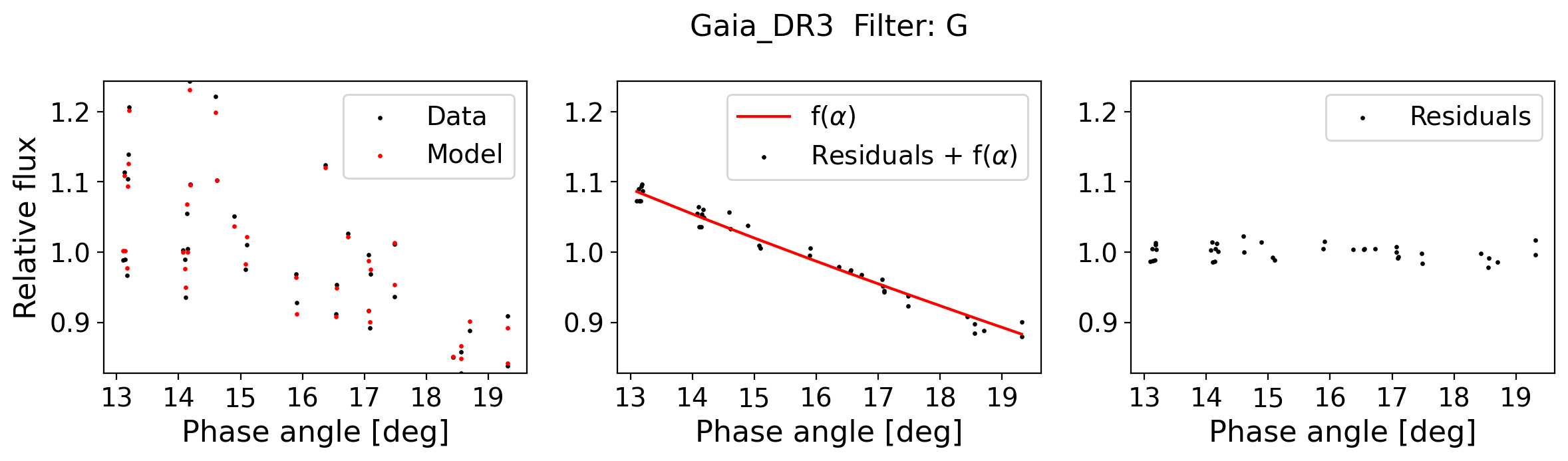}
    \includegraphics[width=0.81\linewidth]{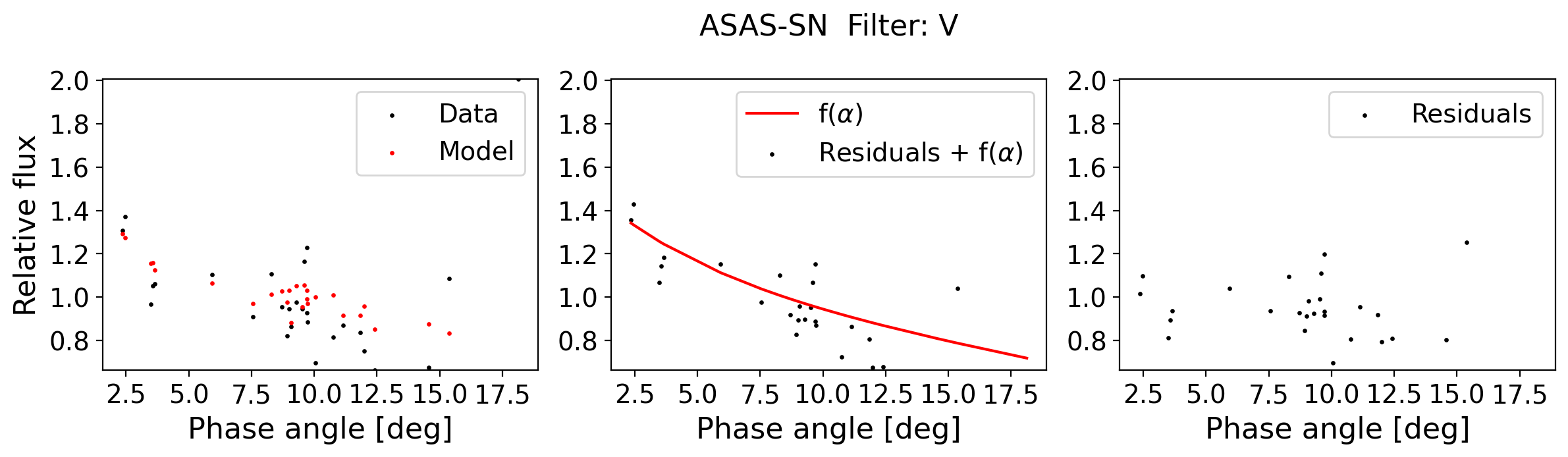}
    \includegraphics[width=0.81\linewidth]{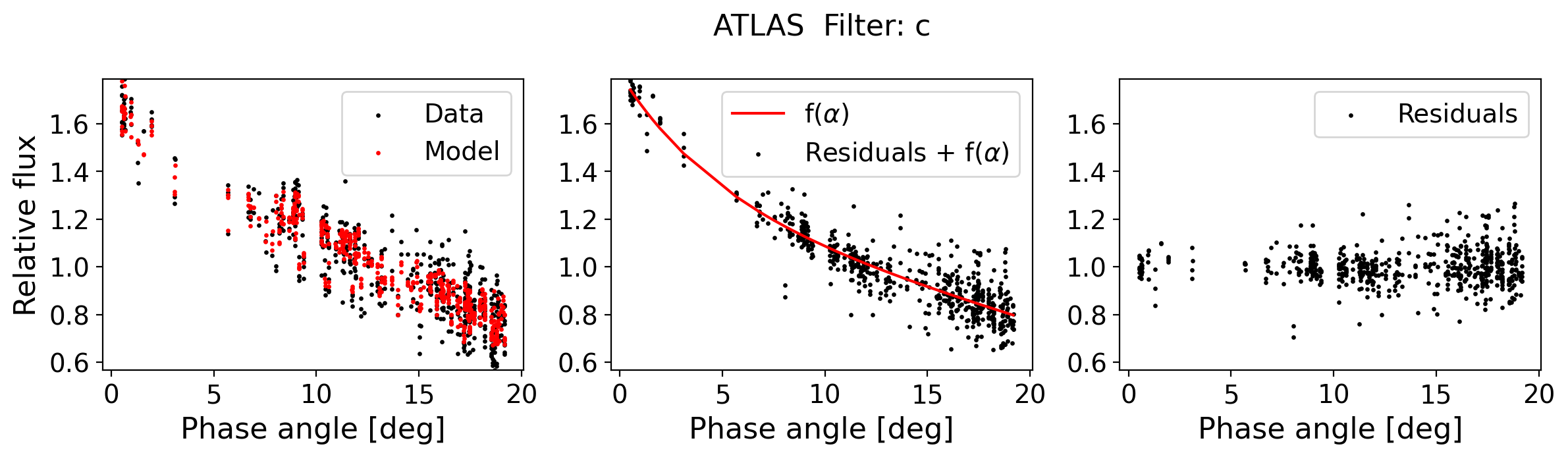}
    \includegraphics[width=0.81\linewidth]{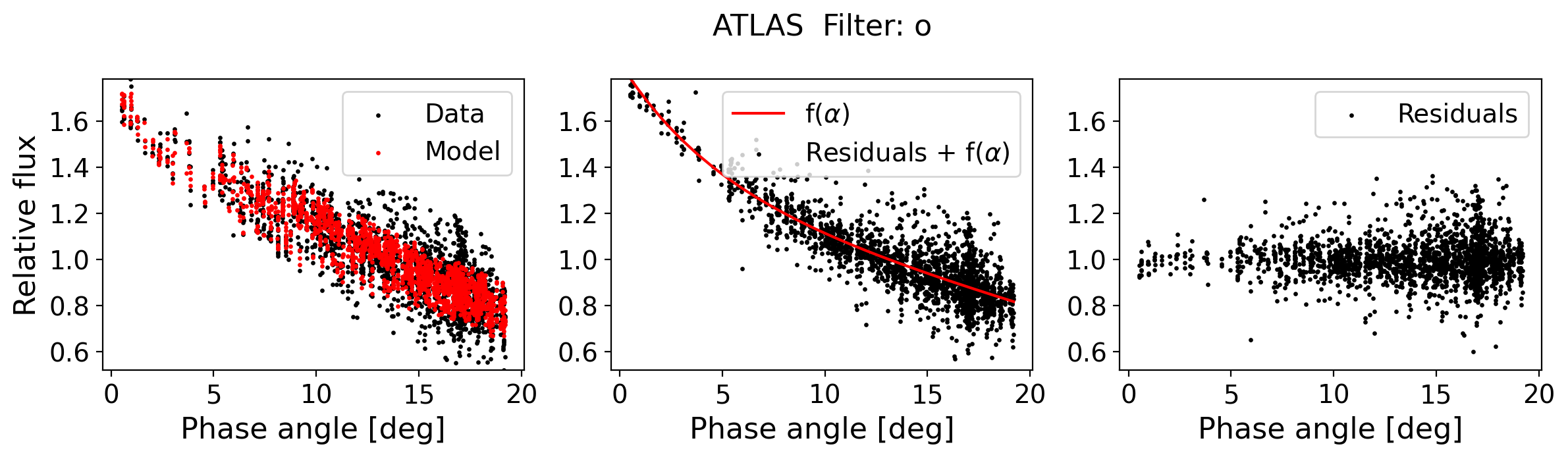}
    \caption{
    Lightcurve fits of asteroid Oersted for sparse photometric datasets from Gaia DR3, ASAS-SN, and ATLAS (c and o filters). 
    For each dataset, three panels are shown: 
    (i) observed and model data, 
    (ii) residuals relative to the fitted phase function, 
    (iii) residuals relative to the normalized lightcurve model. 
    The residuals illustrate the quality of the photometric calibration and the goodness of fit.}
    \label{fig:lcs_fit1}
\end{figure*}

\begin{figure*}
    \centering
    \includegraphics[width=0.81\linewidth]{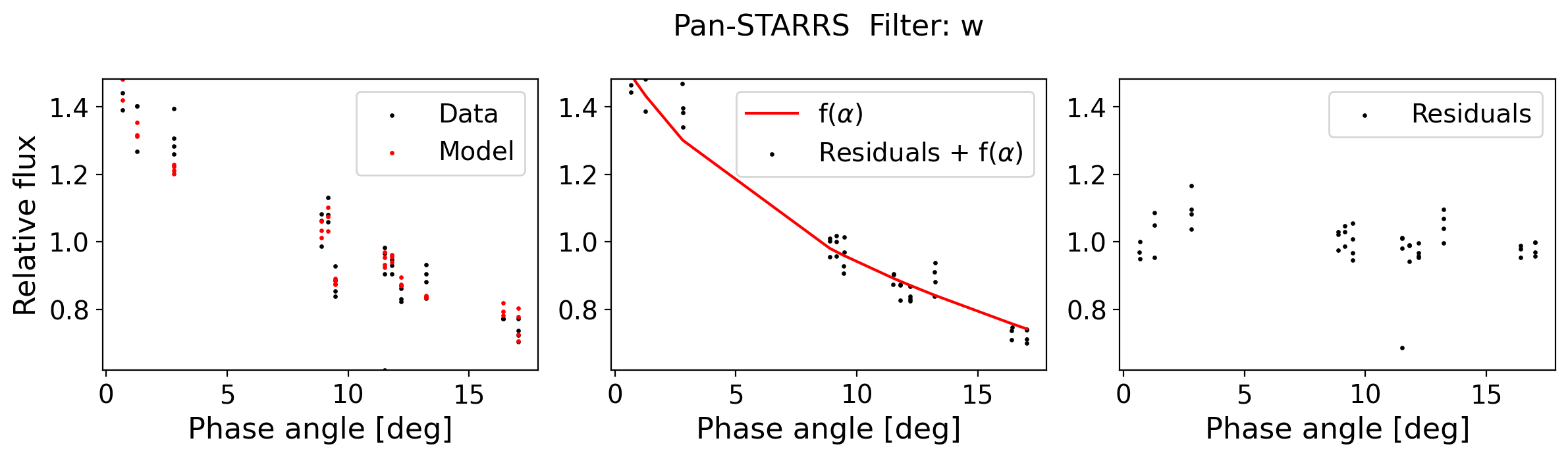}
    \includegraphics[width=0.81\linewidth]{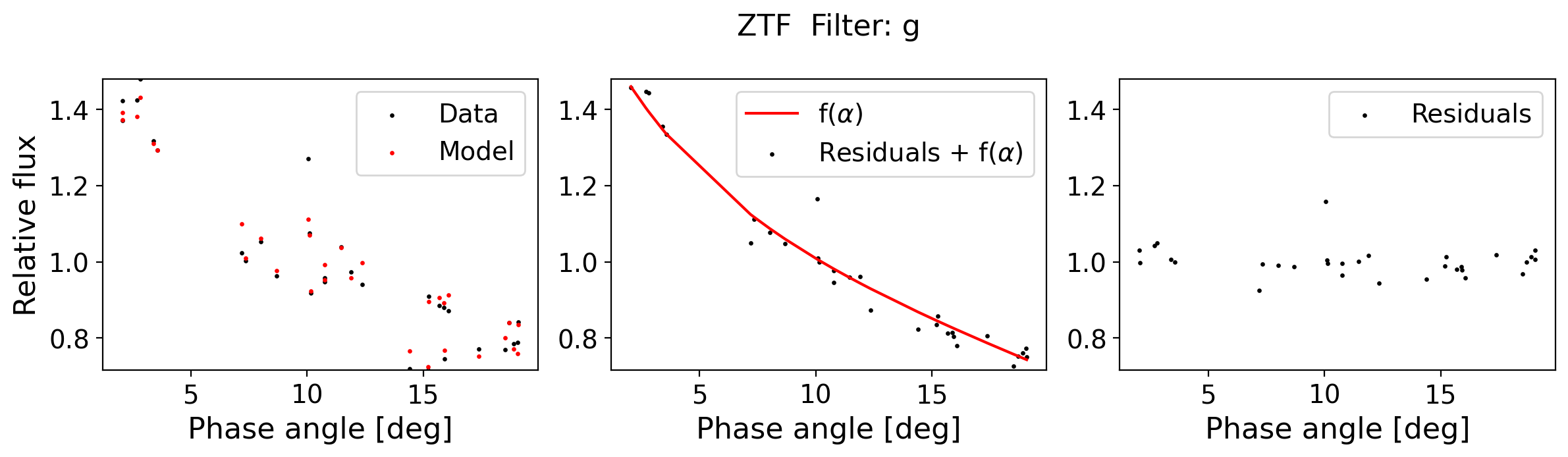}
    \includegraphics[width=0.81\linewidth]{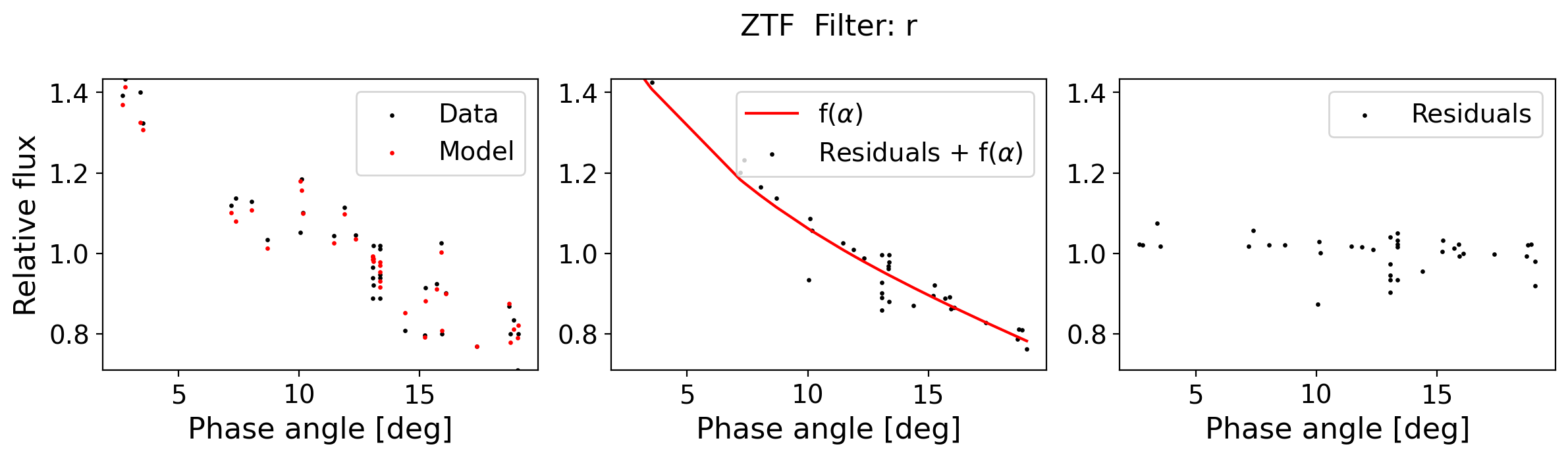}
    \includegraphics[width=0.81\linewidth]{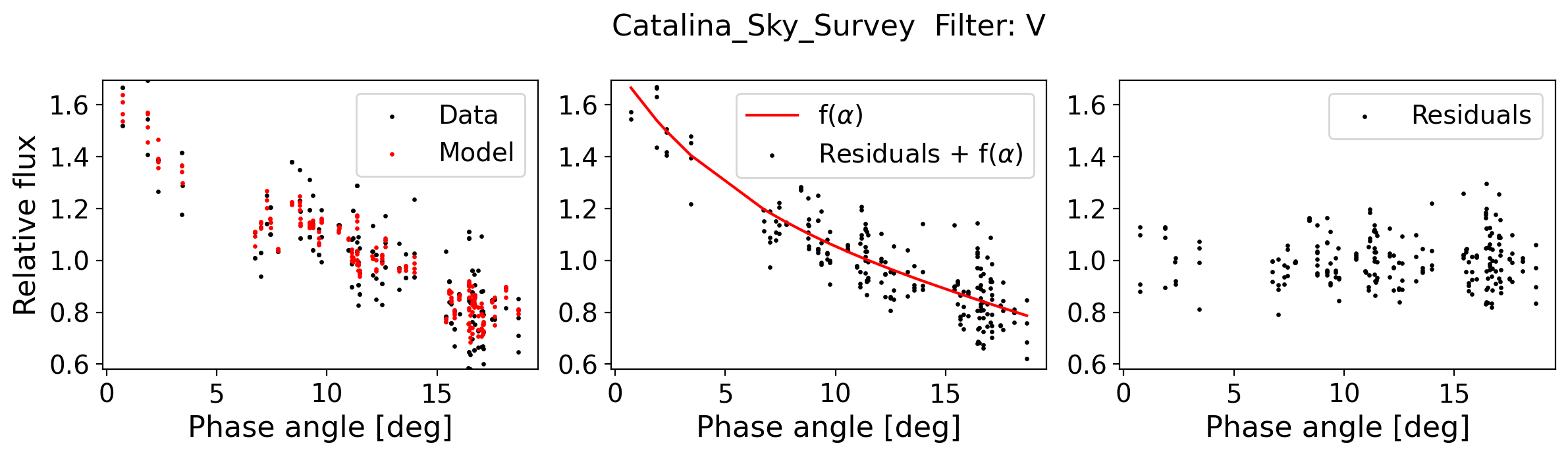}
    \caption{
    Lightcurve fits of asteroid Oersted for sparse photometric datasets from Pan-STARRS, ZTF (g and r filters), and the Catalina Sky Survey. 
    For each dataset, three panels are shown: 
    (i) observed and model data, 
    (ii) residuals relative to the fitted phase function, 
    (iii) residuals relative to the normalized lightcurve model. 
    The residuals illustrate the quality of the photometric calibration and the goodness of fit.}
    \label{fig:lcs_fit2}
\end{figure*}

\end{appendix}

\end{document}